\documentclass[journal]{IEEEtran}
\usepackage{booktabs} % For formal tables
\usepackage{graphicx}
\usepackage{subcaption}
\usepackage{algorithm}
\usepackage{algpseudocode}
\usepackage{diagbox}
\usepackage{rotating}
\usepackage{slashbox,amsmath}
\usepackage{multirow}

\usepackage{tikz}
\usetikzlibrary{fit,shapes.misc}
\newcommand\marktopleft[1]{%
    \tikz[overlay,remember picture] 
        \node (marker-#1-a) at (0,1.5ex) {};%
}
\newcommand\markbottomright[1]{%
    \tikz[overlay,remember picture] 
        \node (marker-#1-b) at (0,0) {};%
    \tikz[overlay,remember picture,thick,solid,inner sep=3pt]
        \node[draw,rounded rectangle,fit=(marker-#1-a.center) (marker-#1-b.center)] {};%
}

\usepackage[markup=underlined]{changes}

\begin{document}

%\title{Attack detection of IoT system based on application characterizing}
\title{Network Phenotyping for Network Traffic Classification and Anomaly Detection}

\author{Minhui~Zou,~\IEEEmembership{Student Member,~IEEE,}
        Chengliang~Wang,~\IEEEmembership{Member,~IEEE,}
        Fangyu~Li,
        and~WenZhan~Song,~\IEEEmembership{Senior Member,~IEEE}% <-this % stops a space
\IEEEcompsocitemizethanks{\IEEEcompsocthanksitem Minhui~Zou and Chengliang~Wang are with the College of Computer Science, Chongqing University, Chongqing, China, 400044. 
Minhui~Zou, Fangyu~Li and WenZhan~Song are with the College of Engineering, University of Georgia, Georgia, US, 30602. 
\protect\\
E-mails: zouminhui@outlook.com, wangcl@cqu.edu.cn; fangyu.li@uga.edu and wsong@uga.edu.
\IEEEcompsocthanksitem Chengliang~Wang is the corresponding author.
\IEEEcompsocthanksitem This work is supported by the National Natural Science Foundation of China under grand No. 61672115 and Chongqing Social Undertakings and Livelihood Security Science and Technology Innovation Project Special Program No. cstc2017shmsA30003..}
% <-this stops a space
\thanks{Manuscript received April 19, 2005; revised August 26, 2015.}}

\markboth{Journal of \LaTeX\ Class Files,~Vol.~14, No.~8, August~2015}%
{Shell \MakeLowercase{\textit{et al.}}: Bare Demo of IEEEtran.cls for IEEE Journals}

\maketitle

\begin{abstract}
%\commentWS{Please pay attention to all my revisions in your abstract and revise your main text accordingly. The current presentation logic is not logical and hard to follow. Do not emphasize the embedded device or remote attestation, as it will mislead people. All your did is just one step toward a big picture.}

This paper proposes to develop a network phenotyping mechanism based on network resource usage analysis and identify abnormal network traffic. 
The network phenotyping may use different metrics in the cyber physical system (CPS), including resource and network usage monitoring, physical state estimation. 
The set of devices will collectively decide a holistic view of the entire system through advanced image processing and machine learning methods. 
In this paper, we choose the network traffic pattern as a study case to demonstrate the effectiveness of the proposed method, while the methodology may similarly apply to classification and anomaly detection based on other resource metrics. 
%\commentWS{You need to summarize the approaches and innovations of the proposed methods here. Do not be fuzzy - be specific. What is new and what is the contribution? For the experiment results, you need to give numbers such as 95\% accuracy.}
% To the best of our knowledge, this work is the first to characterize a whole network system through network resource usage analysis.
% We apply texture feature analysis on the network resource usage of the devices to extract communication patterns.
% Then $k$-NN classification is used to clustering the communication patterns.
We apply image processing and machine learning on the network resource usage to extract and recognize communication patterns.
The phenotype method is experimented on four real-world decentralized applications. 
With proper length of sampled continuous network resource usage, the overall recognition accuracy is about 99\%.
Additionally, the recognition error is used to detect the anomaly network traffic.
We simulate the anomaly network resource usage that equals to 10\%, 20\% and 30\% of the normal network resource usage.
The experiment results show the proposed anomaly detection method is efficient in detecting each intensity of anomaly network resource usage.   
\end{abstract}


\IEEEpeerreviewmaketitle

\section{Introduction}
\IEEEPARstart{W}{ith} the trending of cyber physical system (CPS), computing and connectivity have been present in every aspect of human lives~\cite{2017Gartner2016}. 
Interconnected devices in a network cooperate in the distribution of computing in wireless network, facilitating the decentralized applications, such as seismic monitoring and industry control~\cite{Miorandi2012InternetChallenges}. 
However, due to the restricted computing ability and energy consummation limitation, complicated cryptography could not be implemented on devices of low hardware configuration in CPS. 
Besides, the bugs of software, operating systems and firmwares that running on them lead them to be vulnerable \cite{Arias2017SecurityEra}. 
Malwares like Stunex \cite{Langner2011Stuxnet:Weapon} and Duqu \cite{BencsathB.PekG.ButtyanL.andFelegyhazi2011Duqu:Wild} have caused a lot of concerns. 
Thus, how to ascertain the devices of a CPS are trustworthy is becoming an important and open question.

To mitigate increasing attacks targeting CPS devices, a lot of countermeasures has been proposed. 
Remote attestation is an efficient to verify the security of the individual CPS devices and has been a hot research topic. 
The main idea of remote attestation is the system management sends attest requests to the remote devices and make decisions according to the received security reports from the devices. 
Generally speaking, remote attestation could be sorted into three categories: software-based, hardware-based and hybrid. 
Software-based methods \cite{Seshadri2005Pioneer:Systems} calculate the checksum of memory and essentially register and expect a time-limited response for trustworthy devices without the help of additional hardware. 
Hardware-based methods depend on secure hardware such as TPM \cite{Group-Trusted-Computing2014TPMSpecifications} and TrustZone \cite{Winter2008TrustedPlatforms} to provide secure execution environment. 
The hybrid methods require the minimal collection of hardware and software components that result in secure remote attestation \cite{Francillon2014AAttestation, EldefrawyKarimandTsudikGeneandFrancillonAurelienandPerito2012SMARTRust, Koeberl2014TrustLite:Devices, Brasser2015TyTAN:Devices}. 
However, most existing methods are not scalable for a large number of devices and could only measure static integrity. 

Resource usage analysis is another approach to verify the trustworthiness of CPS devices.
The usage of resources such as CPU, memory and network of the devices is used in resources usage methods.
The inside states of devices could thus be inferred, which is useful for system management. 
\cite{Liu2009VirusMeter:Spies, Kang2011UsageSmartphones} analyzed the user behaviors by tracking the energy consumptions. 
However, they are only targeting individual device. 
\cite{Evans2014ComprehensiveStats, Sorkunlu2017TrackingData} proposed to monitor HPC systems by collecting detailed resource usage information to detect anomalous behaviors. 
But they didn't consider the co-occurrence of resources usage among the HPCs. 
Besides, collecting and transmitting detailed information might not be practical for devices of low hardware configuration which have limited computing and bandwidth resources. 
In a CPS system, interconnected devices work together in running decentralized applications. 
The devices will collectively decide a holistic view of the entire system
Unlike desktop computers or servers, a CPS only runs very limited number of applications simultaneously. 
Thus, the resource usage patterns of a CPS could be characterized by analysis the resource usage patterns of decentralized applications that would run on the CPS devices.
%due to the restricted computing ability and energy consummation limitation. 

Machine learning methods are often combined with resource usage analysis for behavior monitoring of CPS devices \cite{Liu2009VirusMeter:Spies,Evans2014ComprehensiveStats,Sorkunlu2017TrackingData,Caviglione2016SeeingIntelligence}.
Machine learning tasks are typically classified into two broad categories: supervised learning and unsupervised learning.
Supervised learning is presented with example inputs and their desired outputs and the goal is to learn a general rule that maps inputs to outputs.
However, unsupervised learning is given no labels for the inputs and the goal is to discover the hidden patterns of the inputs.
\cite{Liu2009VirusMeter:Spies} used regression, a supervised learning method, to predict the energy consumption of cellphones.
\cite{Caviglione2016SeeingIntelligence} used binary decision tree, also a supervised learning method, to classify the the behaviors of devices.

%\commentFL{the paper is short, maybe you need to add content about machine learning and how temporal information can help in the introduction part.}

In this paper, we choose the network traffic pattern as a study case to demonstrate the effectiveness of the proposed method, while the methodology may similarly apply to classification and anomaly detection based on other resource metrics.
We first apply texture feature analysis to extract the spatial information of the network resource usage of the CPS devices a time point.
Then we study the temporal information of the extracted texture features in time series.
The spatial and temporal features together of the network resource usage determine the communication patterns of the CPS network.
At last, a supervised learning method, $k$-NN classification, is used to classify and recognize the communication patterns.
The phenotype method is experimented on four real-world decentralized applications. 
With proper length of sampled continuous network resource usage, the overall classification accuracy is about 99\%.
Additionally, the classification error is used to detect the anomaly network resource usage.
We simulate the anomaly network resource usage that equals to 10\%, 20\% and 30\% of the normal network resource usage.
The experiment results show the proposed anomaly detection method is efficient in detecting each intensity of anomaly network resource usage.

The contributions of this work are summarized below:
\begin{itemize}
\item To our best knowledge, this work is the first to phenotype a whole CPS network instead of individual devices.
\item This work presents a novel method to characterize the network resource usage of four real-work decentralized applications based on image processing and machine learning. 
\item This work also propose a straightforward and efficient method to detect abnormal network resource usage of CPS network. 
\end{itemize}

The rest of this paper is structured as follows. 
Section \ref{sec:preliminaries} shows the preliminaries of the proposed approaches. 
Section \ref{sec:Characterizing_the_decentralized_applications_based_on_network_traffic} proposes the method to phenotype the CPS network base on network resource usage analysis and the method to detect anomaly network traffic. 
Section \ref{sec:ExperimentResultsAndDiscussion} shows and discusses the experiment results.
Section \ref{sec:conclusion} concludes this paper.

\section{Preliminaries}
\label{sec:preliminaries}

%\commentFL{why do you call this part preliminaries?}
%\commentMZ{because this whole section is to introduce the system model, background and definitions}

\subsection{System Model}
\label{sec:system_model}

\begin{figure}%[H]
\centering
\includegraphics[width=0.45\textwidth]{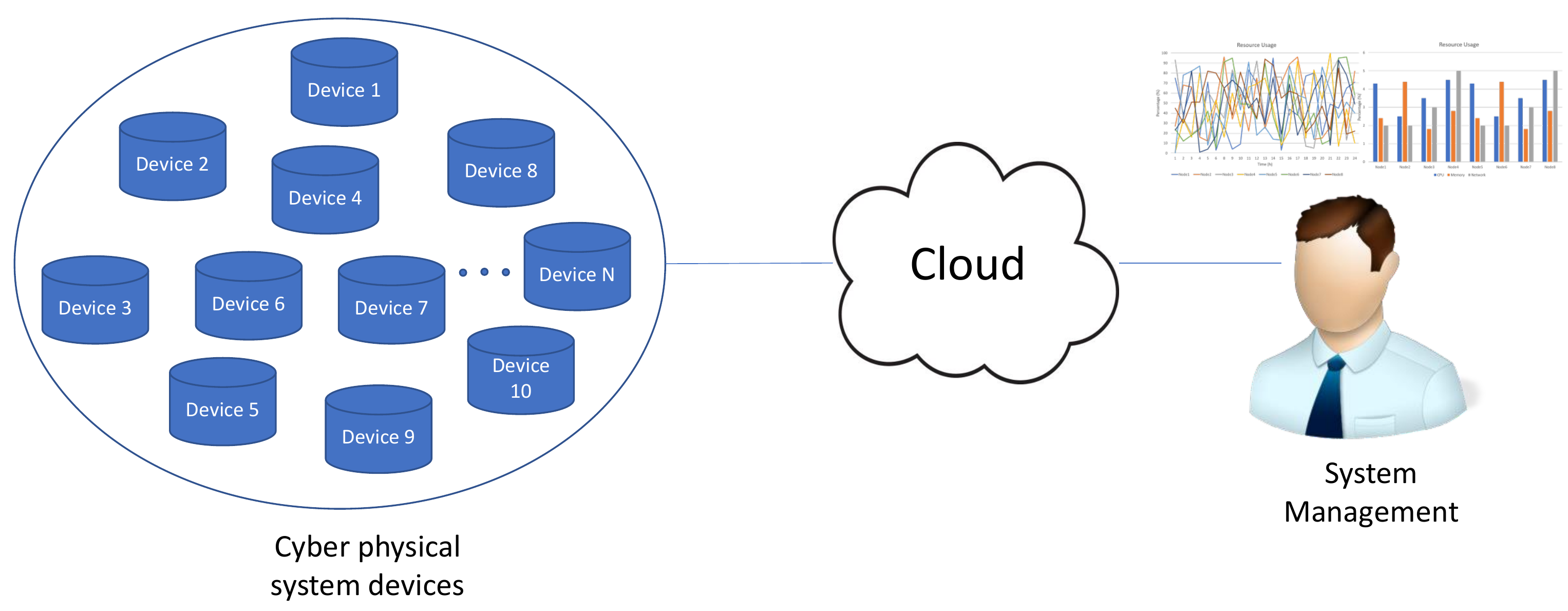}
%\vspace{-0.3cm}
\caption{Cyber physical system}
\label{fig:IoT_System}
\end{figure}

Many CPS use a cloud to visualize and analyze data collected from the CPS devices, as shown in Fig. \ref{fig:IoT_System}.  
System management is provided with information about the states of all the CPS devices for security operation, such as shutting down or restarting the CPS network because of the detected abnormal behaviors. 

In this model, the system management is provided with real-time network resource usage information through the cloud.
From the perspective of sensors, the network resource usage application or process locally running on every CPS device works as a sensor, which monitoring the network resource usage of it.
The set of sensors collectively decide a holistic view of the whole CPS network.
The CPS devices might be equipped with heterogeneous hardware and software configuration and those devices are interconnected.
However, generally, those devices are resource-constrained and only very limited number of applications run on the CPS at the same time.
The applications running on the CPS are decentralized and the devices are scheduled to collect data from their built-in sensors or transfer/receive data for their neighboring devices.
According to the different applications running on the CPS, the CPS devices present application-specific communication feature in when to transfer/receive data and the size of data that are transferred/received.
Hence, we propose to phenotype the CPS network by analyze the communication feature of the applications that would run on the CPS.

\begin{figure}%[H]
\centering
\includegraphics[width=0.35\textwidth]{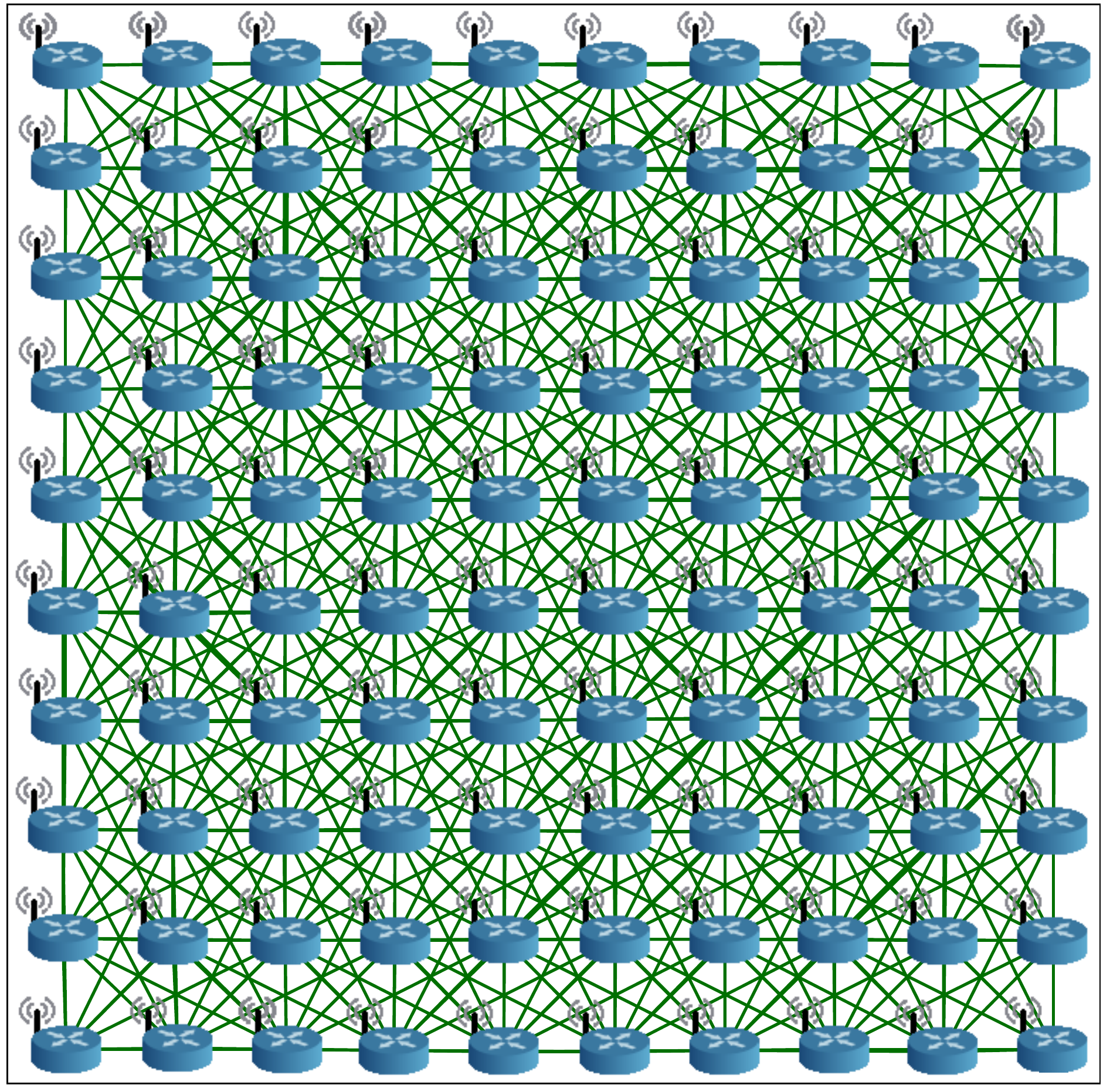}
%\vspace{-0.3cm}
\caption{An example cyber physical system network consisting of 100 devices}
\label{fig:IoT_System_100Nodes}
\end{figure}

As as Fig. \ref{fig:IoT_System_100Nodes}, there are 100 devices in an example CPS network.
Those devices are interconnected with each other directly or indirectly wirelessly.
The discussion of the rest of the paper is based on this example CPS network.
Note that the proposed phenotype method and anomaly detection method is not depend on the topology of the CPS network.

\subsection{Threat Model}
\label{sec:attack_model}
Denial of service (DoS) attack and botnet attack are two common attacks against CPS network \cite{Stavrou2017DDoSIoT}.
A DoS attack could even drain the limited battery life and computational capabilities of the CPS devices, which could cause them to be slow to response to other devices. 
A heavy DoS attack could lead the devices to be completely irresponsible, which is trivial to detect.
A botnet attack is to control the CPS devices to carry out distributed Dos attacks to remote target servers.
An aggressive botnet attack could consume a lot of network resource and cause the applications running on the CPS network to progress very slowly or halt, which is also conspicuous to the system management.
Hence, in this paper, we consider a more sophisticated attack that brings in additional network traffic among the CPS devices but would not affect the progress of the applications running on the CPS network.
The attack could be either a light DoS attack or light botnet attack.
Both the attacks introduce anomaly network traffic to the CPS network.

\subsection {Definitions}
For the ease of discussion, let's introduce the definitions for this paper.

\newtheorem{name}{Printed output}
\newtheorem{mydef}{Definition}
\begin{mydef}
$\textbf{Network throughput}$, the real-time sum of speeds of transferring and sending data of a CPS device;
\end{mydef}

\begin{mydef}
$\textbf{Communication matrix}$, a matrix comprised of the network throughput of all the CPS device at a time;
\end{mydef}

\begin{mydef}
$\textbf{Communication picture}$, a picture created by visualizing a communication picture;
\end{mydef}

\begin{mydef}
$\textbf{Texture feature snippet}$, a slice of the time-series texture features;
\end{mydef}

\begin{mydef}
$\textbf{Coefficient matrix}$, each element of which is the Pearson Correlation Coefficient between two time-series texture feature of a texture feature snippet.
\end{mydef}

\begin{mydef}
$\textbf{Communication pattern}$, partial elements of a coefficient matrix.
\end{mydef}

\section{Phenotyping the CPS network and detecting anomaly network traffic}
\label{sec:Characterizing_the_decentralized_applications_based_on_network_traffic}
Fig. \ref{fig:IoT_System_100Nodes} shows that the 100 interconnected devices in the example CPS network.
The communication between every device and other devices is scheduled by the decentralized applications running on the CPS network.
Each application has its own unique schedule.
Based on this observation, we propose to phenotype the CPS network by analyzing the network traffic of the  decentralized applications running on it.

\subsection{Spatial analysis of the communication image}
\begin{figure}%[H]
\centering
\includegraphics[width=0.48\textwidth]{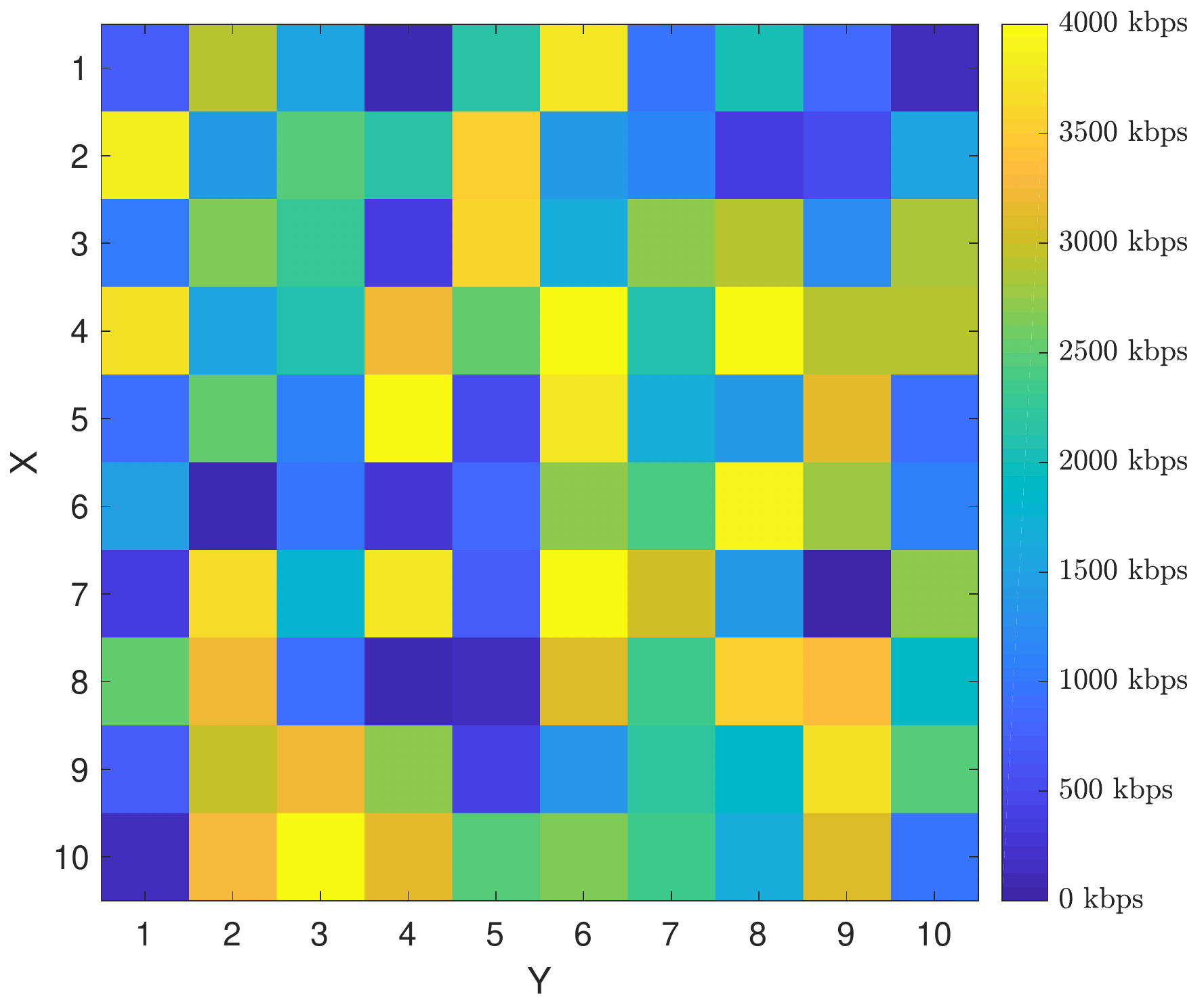}
%\vspace{-0.5cm}
%\caption{A communication image of 100 nodes}
\caption{An example communication image of the example cyber physical system network}
\label{fig:A_communication_image_of_100_nodes}
\end{figure}

By reading the network throughput of every device in Fig. \ref{fig:IoT_System_100Nodes} and organize them as the devices are ordered in the figure, we get an communication image by visualizing the network throughput of every device as a pixel.
As shown in Fig. \ref{fig:A_communication_image_of_100_nodes}, an example communication image of CPS network in Fig. \ref{fig:IoT_System_100Nodes}. 
There are 100 pixels in the figure and $x$ stands for the $x$th row of the picture and $y$ stands for the $y$th column of the picture.
$p(x,y)$, the value of the $x$th row and $y$th column pixel, is the network throughput of the $(10*x+y)$th device.
Note that if the topology of the CPS network is squared as that in Fig. \ref{fig:IoT_System_100Nodes}, we could still get a rectangular communication picture by padding with some zero-value pixels, which would not affect the following proposed methods.

Each communication image represents the network resource usage of the whole CPS network at a time sample.
According to the different applications running on it at that time, the corresponding communication images are different.
In this paper, we utilize texture feature analysis \cite{2002TextureExtraction,Materka1998TextureReview,Zhao2016CharacterizingMaps} on the communication images. 
Texture analysis is an important element of human vision and has been apply to problems of texture classification, discrimination and segmentation \cite{2002TextureExtraction}.
According to \cite{Baraldi1995InvestigationParameters}, there are many different kinds of texture features. 

In this paper, we propose to transform the communication image into multiple single values to reduce computing complexity. 
Hence we choose the statistics based feature extraction methods. 
Statistics based feature extraction methods are classified into three classes: first-order statistics, second-order statistics and higher-order statistics. 
First-order statistics only consider the individual pixels. 
Second-order statistics depend on the co-occurrence of neighboring pixels. 
Higher-order statistics are to deal with nonlinearities and non-Gaussianity. 
In our case, the interconnected devices of the CPS network work collaboratively so that the network throughput of neighboring devices are related to each other. 
Hence, this paper chooses GLCM (gray level co-occurrence matrix) texture features, the most popular second-order statistical texture features. 
We select five GLCM texture features: energy, entropy, contrast, IDM (Inverse Difference Moment) and DM (Directional Moment) \cite{Baraldi1995InvestigationParameters}. 
To get the GLCM texture features, we need to first convert the communication images into GLCM.

\subsubsection{Convert the communication images into GLCM}
Shown as in Table \ref{tab:Intensity_values_of_example_communication_image} are the corresponding intensity (gray-level) values of every pixel of Fig. \ref{fig:IoT_System_100Nodes}.
Note that we normalize the network throughput of the devices to be within the range of $0$ to $4$ for the ease of demonstrating how to convert the communication image into GLCM.
Let's denote the length of the range as $L$ and in this example $L$ equals to $5$.

\begin{figure}[h!]
    \captionsetup[subfigure]{position=b}
    \centering
    \subcaptionbox
        {Horizontal \label{fig:GLCM_adjacent_relationship_horizontal}}
        {\includegraphics[width=.25\linewidth]{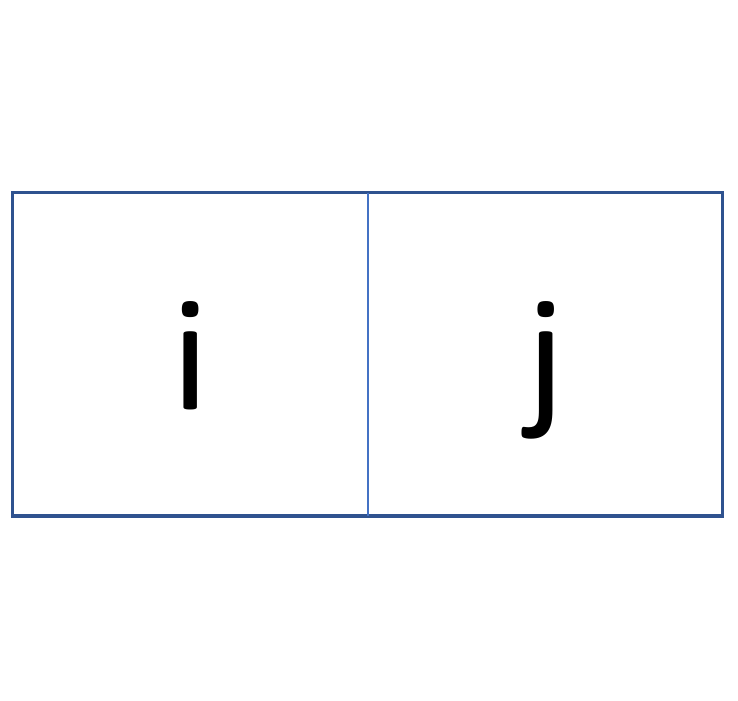}}
    \subcaptionbox
        {Vertical \label{fig:GLCM_adjacent_relationship_vertical}}
        {\includegraphics[width=.25\linewidth]{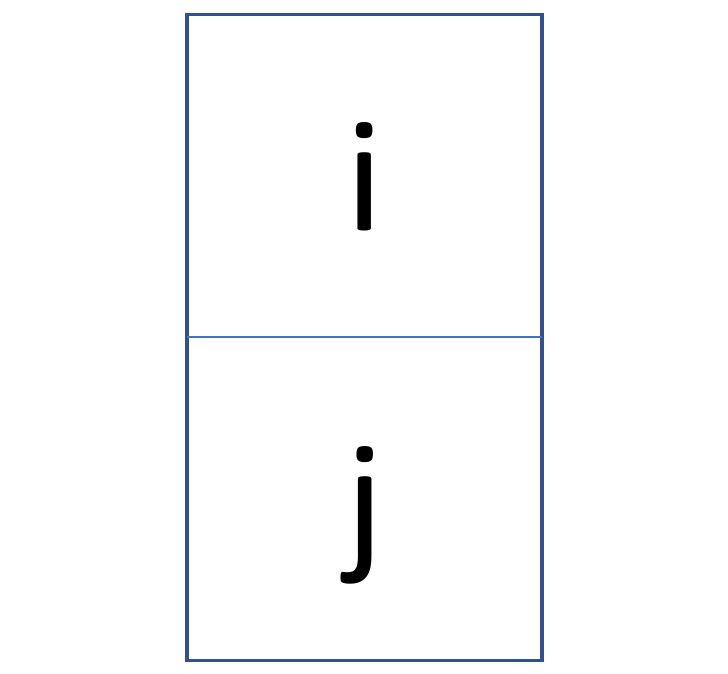}}
    \\
    \subcaptionbox
        {Positive diagonal \label{fig:GLCM_adjacent_relationship_positive_diagnal}}
        {\includegraphics[width=.25\linewidth]{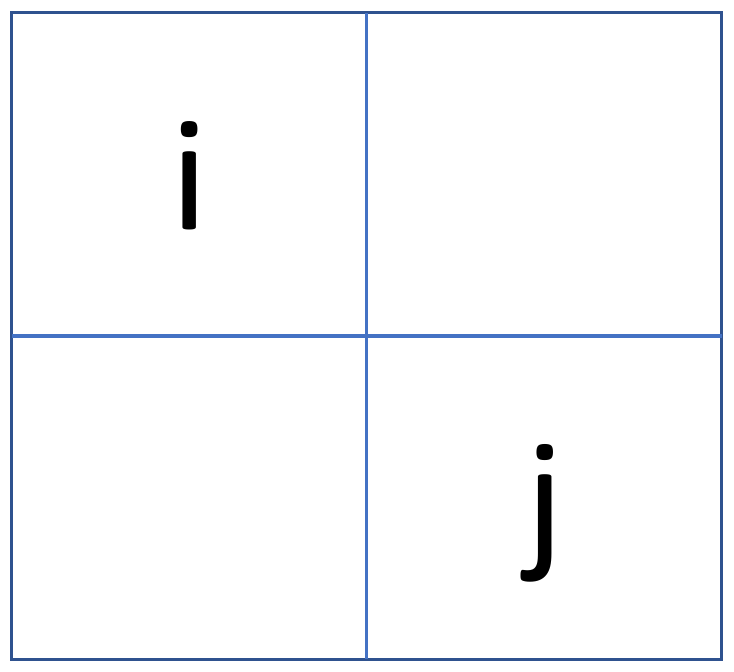}}
    \subcaptionbox
        {Anti-diagonal \label{fig:GLCM_adjacent_relationship_anti_diagnal}}
        {\includegraphics[width=.25\linewidth]{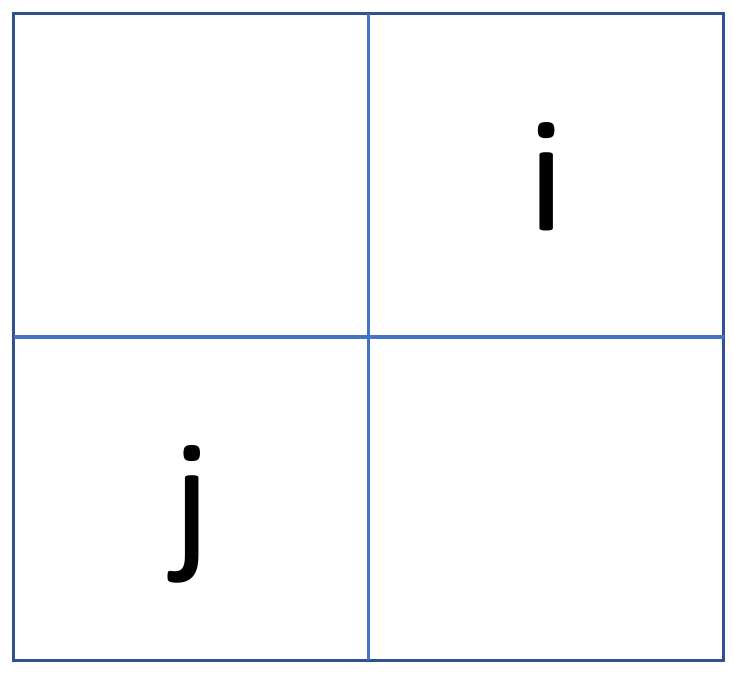}}
    \caption{Four different adjacent relationship between two pixels with value $i$ and $j$, respectively.
    %\commentFL{can you make c and d in the second row?}
    }
    \label{fig:Four_different_adjacent_relationship}
\end{figure}

Table \ref{tab:GLCM_of_example_communication_image} is the resultant GLCM.
$i$ and $j$ stand for the $i$th row and $j$th column of the GLCM matrix, respectively.
Each element $\#(i,j)$ in the GLCM is the sum of the number of times that the pixel with value $i$ occurred in the specified spatial relationship to a pixel with value $j$ in the communication image \cite{Haralick1973TexturalClassification}.
There are four different kinds of relationship between two adjacent pixels: horizontal, vertical, positive diagonal and anti-diagonal, as shown in Fig. \ref{fig:GLCM_adjacent_relationship_horizontal}, \ref{fig:GLCM_adjacent_relationship_vertical}, \ref{fig:GLCM_adjacent_relationship_positive_diagnal} and \ref{fig:GLCM_adjacent_relationship_anti_diagnal}, respectively.
Take the horizontal relationship for example: as shown in Table \ref{tab:Intensity_values_of_example_communication_image}, all the horizontal adjacent pixel pairs that is (1,2) are circled.
Hence it is easy to see $\#(1,2)$, circled in Table Table \ref{tab:GLCM_of_example_communication_image}, equals to 6. 

\begin{table}[H]
    \centering
    \caption{Intensity values of the example communication image.}
    \label{tab:Intensity_values_of_example_communication_image}
    \scalebox{1.2}{
        \begin{tabular}{|c|c|c|c|c|c|c|c|c|c|}
          \hline
          1     &3     &2     &0     &2     &4     &\marktopleft{c1}1     &2\markbottomright{c1}     &1     &0 \\ \hline
     	  4     &\marktopleft{c2}1     &2\markbottomright{c2}     &2     &4     &1     &1     &0     &\marktopleft{c3}1     &2\markbottomright{c3} \\ \hline
          1     &3     &2     &0     &4     &2     &3     &3     &1     &3 \\ \hline
          4     &2     &2     &3     &3     &4     &2     &4     &3     &3 \\ \hline
          1     &3     &1     &4     &1     &4     &2     &1     &3     &1 \\ \hline
          1     &0     &1     &0     &1     &3     &2     &4     &3     &1 \\ \hline
          0     &4     &2     &4     &1     &4     &3     &1     &0     &3 \\ \hline
          3     &3     &1     &0     &0     &3     &2     &4     &3     &2 \\ \hline
          1     &3     &3     &3     &0     &\marktopleft{c4}1     &2\markbottomright{c4}     &2     &4     &2 \\ \hline
          0     &3     &4     &3     &2     &3     &2     &2     &3     &1 \\ \hline
          \marktopleft{c5}1     &2\markbottomright{c5}     &0     &3     &2     &\marktopleft{c6}1     &2\markbottomright{c6}     &1     &3     &4 \\ \hline
        \end{tabular}
    }
\end{table}

% \begin{table}[H]
%     \centering
%     \caption{GLCM (co-occurrence matrix) of the example communication image}
% 	\label{tab:GLCM_of_example_communication_image}
%     \scalebox{1.2}{
%         \begin{tabular}{|c|c|c|c|c|c|}
%           \hline
%           \#(0,0)    & \#(0,1)    & \#(0,2)   & \#(0,3)   & \#(0,4) \\ \hline
%           \#(1,0)    & \#(1,1)    & \#(1,2)   & \#(1,3)   & \#(1,4) \\ \hline
%           \#(2,0)    & \#(2,1)    & \#(2,2)   & \#(2,3)   & \#(2,4) \\ \hline
%           \#(3,0)    & \#(3,1)    & \#(3,2)   & \#(3,3)   & \#(3,4) \\ \hline
%           \#(4,0)    & \#(4,1)    & \#(4,2)   & \#(4,3)   & \#(4,4) \\ \hline
%         \end{tabular}
%     }
% \end{table}

\begin{table}[H]
    \centering
    \caption{GLCM (co-occurrence matrix) of the example communication image}
	\label{tab:GLCM_of_example_communication_image}
    \scalebox{1.2}{
        \begin{tabular}{l|ccccc}
			\backslashbox{i}{j} & 0          & 1          & 2         & 3         & 4       \\ \hline
			0                   & \#(0,0)    & \#(0,1)    & \#(0,2)   & \#(0,3)   & \#(0,4) \\ %\cline{2-6}
            1                   & \#(1,0)    & \#(1,1)    & \marktopleft{c11}\#(1,2)\markbottomright{c11}   & \#(1,3)   & \#(1,4) \\ %\cline{2-6}
            2                   & \#(2,0)    & \#(2,1)    & \#(2,2)   & \#(2,3)   & \#(2,4) \\ %\cline{2-6}
            3                   & \#(3,0)    & \#(3,1)    & \#(3,2)   & \#(3,3)   & \#(3,4) \\ %\cline{2-6}
            4                   & \#(4,0)    & \#(4,1)    & \#(4,2)   & \#(4,3)   & \#(4,4) \\ %\cline{2-6}
		\end{tabular}
    }
\end{table}

%Calculating how often a pixel with the intensity (gray-level) value $i$ occurs in a specific spatial relationship to a pixel with the value $j$.  
As we can see, a GLCM is a $L$ by $L$ matrix. 
The larger $L$ is, more information of the communication picture is preserved in the GLCM.
However, larger $L$ brings in more computing complexity.
%\commentMZ{Note that the size of the GLCM $L$ does not depend on the number of devices in the CPS network.} 

\subsubsection{Calculate GLCM texture features}
As discussed in previous section, we can get four GLCMs from a communication picture from the perspectives of four different adjacent relationship between two pixels in the communication picture.
For each GLCM, we apply below five equations \cite{Baraldi1995InvestigationParameters}.

Energy:
\begin{equation}
Energy=\sqrt{\sum_{i=1}^{L}\sum_{j=1}^{L}M^{2}(i,j)}.
\end{equation}

Entropy:
\begin{equation}
Entropy=\sum_{i=1}^{L}\sum_{j=1}^{L}M(i,j)(-\ln(M(i,j))).
\end{equation}

Contrast:
\begin{equation}
Contrast=\sum_{i=1}^{L}\sum_{j=1}^{L}(i-j)^{^{2}}M(i,j).
\end{equation}

IDM (Inverse Difference Moment):
\begin{equation}
IDM=\sum_{i=1}^{L}\sum_{j=1}^{L}\frac{1}{1+(i-j)^2}M(i,j).
\end{equation}

DM (Directional Moment): 
\begin{equation}
DM=\sum_{i=1}^{L}\sum_{j=1}^{L}\left|i-j\right|M(i,j).
\end{equation}

Energy, entropy, contrast, IDM and DM measure the intensity of pixel pair repetitions, the randomness of pixels, the extent of a pixel and its neighbors, local homogeneity of the communication image and alignment of it in terms of the angle, respectively.

Let's denote the number of GLCM texture features $N_f$ we get from one communication picture.
For each GLCM from the perspective of one adjacent relationship between two pixel, we get $5$ GLCM texture features.
Thus, we get $5*4=20$ GLCM texture features in total, i.e. $N_f=20$.

\subsection{Temporal analysis of GLCM texture features}
We have previously mentioned that each communication image stands for the network resource usage of the CPS network at a time sample.
The GLCM texture features of a single communication are not enough to represent the application running on the CPS devices.
We consider the communication feature as the transition of the texture features, i.e., the communicate feature is showed by both the $N_{f}$ texture feature values and the time-series trends of them.

\begin{figure}%[H]
\centering
\includegraphics[width=0.53\textwidth]{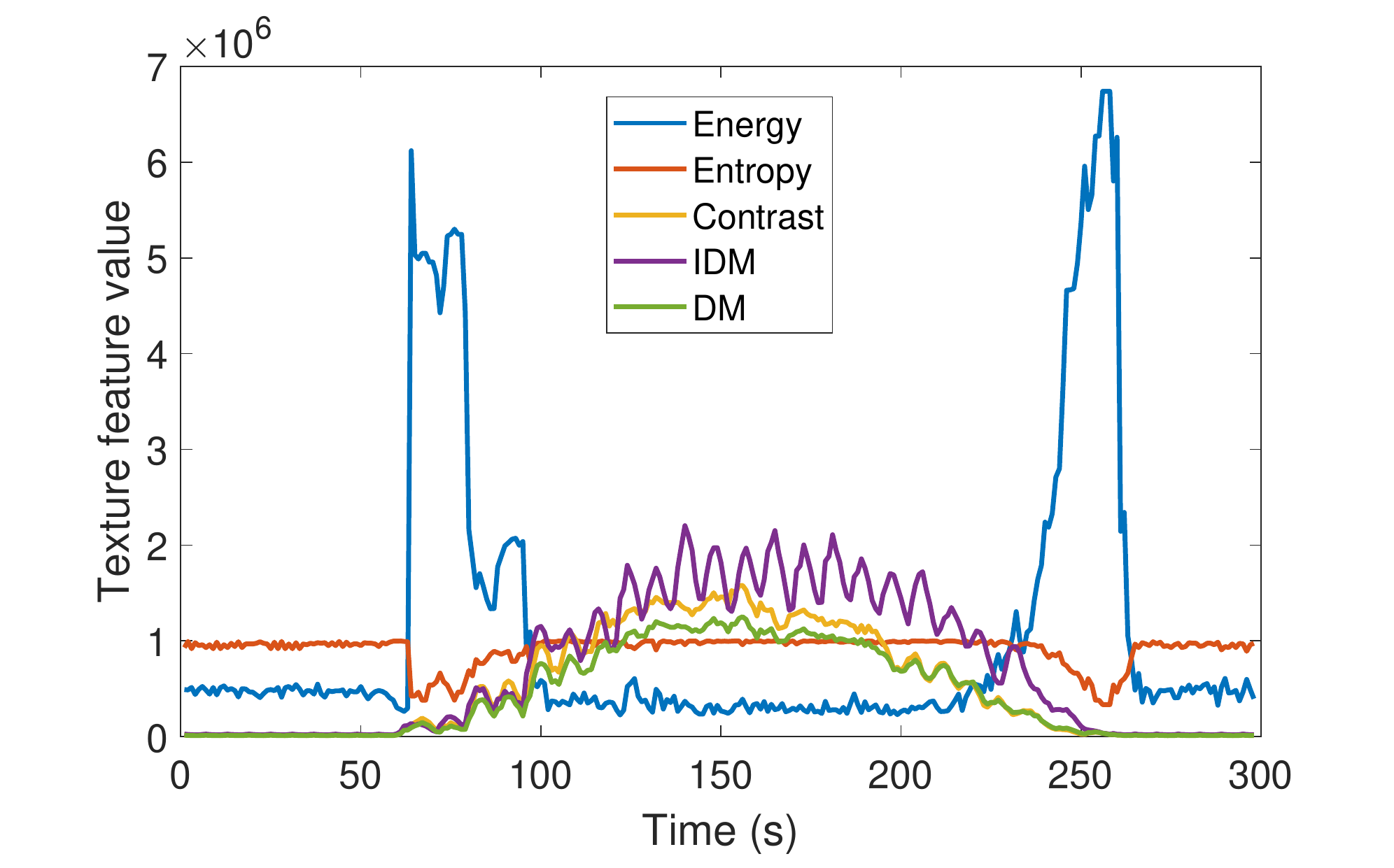}
%\vspace{-0.5cm}
\caption{Time-series trends of texture feature values}
\label{fig:FiveFeaturesOFConsensus}
\end{figure}

As shown in Fig. \ref{fig:FiveFeaturesOFConsensus}, we present the time-series trends of the texture feature values of an example application for only one adjacent relationship between two pixels \footnote{In order to show all the five curves in one picture, we normalize the Energy and Entropy by timing them with $10^5$.}. 
The horizontal axis is time and the vertical axis is texture feature value.
We record the example application every second \footnote{The communication images are recorded every second throughout this paper.} for 300 seconds, starting about 50 seconds before the application starts and around 50 seconds after it ends.
The example application lasts around 200 seconds.
The value of the points the $N_{f}$ curves go through and the time-series trends of the curves representing the communication features of the example application.

\subsubsection{Texture feature snippets} 
\begin{figure}%[H]
\centering
\includegraphics[width=0.50\textwidth]{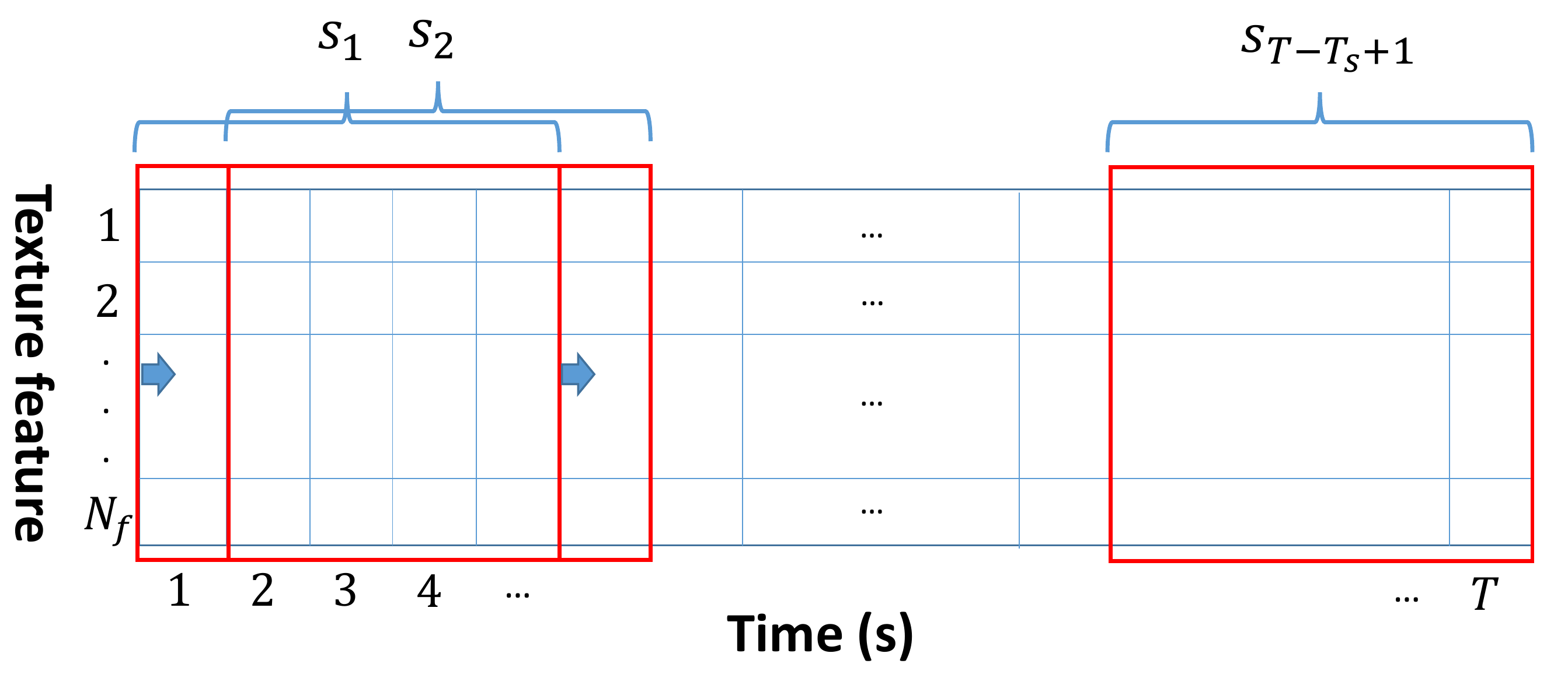}
\caption{Slicing the time-series texture feature trends into texture feature snippets}
\label{fig:Communication_pattern_slicing}
\end{figure}

In order to extract more communication features of one decentralized application ruining on the interconnected CPS network, we slice the time-series texture feature trends into texture feature snippets by using a sliding window, as shown in Fig. \ref{fig:Communication_pattern_slicing}.

Assume the running time of an application is $T$ seconds and the communication images are recorded every second.  
Assume the size of the sliding window is $T_{S}$, then every texture feature snippet consists of $N_{f}*T_{S}$ texture feature values.
The sliding window move once a second, hence we have $T-T_{S}+1$ snippets for this application.

\subsubsection{Coefficient matrix}
We observe that each time the CPS network runs a certain application, all the time-series texture feature trends are stochastic.
However, the similarities between each two time-series texture feature trends are stable.
Hence we propose to use the similarities between each two time-series texture feature trends as the communication patterns.
Pearson Correlation Coefficient \cite{Benesty2009PearsonCoefficient} is common method to calculate the similarity of two vectors.
As shown in Equation \ref{eq:pearson_correlation_coefficient}, $x$ and $y$ are two time-series texture features, $\rho(x,y)$ is the Pearson Correlation Coefficient between $x$ and $y$.

\begin{equation}
\label{eq:pearson_correlation_coefficient}
\rho(x,y) = \frac{n\sum\limits^{N}_{i=1}{x_i y_i}-\sum\limits^{N}_{i=1}{x_i}\sum\limits^{N}_{i=1}{y_i}}
{\sqrt{(n\sum\limits^{N}_{i=1}{x_i}^2-{\sum\limits^{N}_{i=1}{x_i}}^2)(n\sum\limits^{N}_{i=1}{y_i}^2-{\sum\limits^{N}_{i=1}{y_i}}^2)}}
\end{equation}

Let's denote the $i$th time-series texture feature of the texture feature snippet as $v_i$.
By computing the Pearson Correlation Coefficient between every two time-series texture feature trends, we could get a $N_f$ times $N_f$ coefficient matrix as below:

\[
\begin{bmatrix}
    \rho(f_1,f_1)     & \rho(f_1,f_2)     & \rho(f_1,f_3)     & ... & \rho(f_1,f_{N_f}) \\
    \rho(f_2,f_1)     & \rho(f_2,f_2)     & \rho(f_2,f_3)     & ... & \rho(f_2,f_{N_f}) \\
    ...               & ...               & ...               & ... & ...               \\
    \rho(f_{N_f},f_1) & \rho(f_{N_f},f_2) & \rho(f_{N_f},f_3) & ... & \rho(f_{N_f},f_{N_f}) \\
\end{bmatrix}
\]

The coefficient matrix is a symmetric matrix and all of the elements in the diagonal line, $\rho(f_i,f_i)$ ($i\in{[1, N_f]}$), equal 1.
Thus we could use partial elements $\rho(f_i,f_j)$ ($i, j\in{[1, N_f]}$ and $i \neq j$) to represent the coefficient matrix.
The size-reduced coefficient matrix is comprised of only $N_f*(N_f/2-1)$ elements.
We represent the size-reduced coefficient matrix as the communication pattern.
For each application, we have $T-T_{S}+1$ communication patterns.
%Time complexity is $(T-T_{S}+1)*$
Note that computing the communication patterns from the communication images does not depend on the number of devices in the CPS network. 
Hence \textbf{the proposed phenotyping method is scalable}.

\subsection{Classification of the communication patterns}
\subsubsection{Communication pattern labeling}
We denote $N_{A}$ as the number of different applications that would run on the CPS network. 
For application $A_{p} (p\in[1,N_{A}])$, we transform the network throughput into communication patterns, $s_{pq}$ ($q\in[1,T_p-T_{S}+1]$ ), and label them as $L_{p}$.
In this paper, we use $k$-NN ($k$-nearest neighbors algorithm) to classify the communication patterns.
$k$-NN is a very popular supervised machine-learning classification method \cite{Altman1992AnRegression}.
In order to further reduce the computing complexity, we perform principal components analysis on the attributes before training the $k$-NN model.
%To prevent overfitting, we utilize 5-fold crossvalidation. 

\subsubsection{Recognition of Communication Patterns}
Algorithm \ref{alg:Algorithm_Recognition} shows how to recognize the communication patterns.
Firstly, the algorithm is given a communication pattern under recognition, $s_{test}$.
Then it searches the labeled communication patterns to find $k$ communication patterns with shortest Euclidean distance with $s_{test}$ and store them in set $Set_{K}$.
The next step is find the label $L_{p}$ that labels most communication patterns in $Set_{K}$.
Label $L_{p}$ is the result of recognition. 

%\commentFL{Please check the formatting of the algorithm, some steps are not right}
\begin{algorithm}[H]
  \caption{\small{Recognition of a given communication pattern}}
  \label{alg:Algorithm_Recognition}
  \begin{algorithmic}[1]
    \State {Inputs: a communication pattern $s_{test}$ under recognition;}
    \State {Outputs: The label of application the $s_{test}$ belongs to;}
    \State {Set $Set_{K}=\{\}$;}
    \For{each communication pattern $s_{pq}$ in the database;}
        \State {Compute the euclidean distance $d_{pq}$ between $s_{test}$ and $s_{pq}$;}
    \EndFor
    \State {Choose $k$ communication patterns with smallest Euclidean distance with $s_{test}$ and put them in $Set_{K}$;}
    \State {Return the label $L_{p}$ with most communication patterns that labeled as $L_{p}$ in $Set_{K}$;}
  \end{algorithmic}
\end{algorithm}

\subsection{Anomaly detection}
\label{sec:remote_attestation}
Training the $k$-NN model must be done under safe environment, i.e., all the CPS devices are trustworthy. 
The anomaly detection period could be done under unsafe environment.
Recollect the threat model in \ref{sec:attack_model} that the attack would not affect the progress of the applications running on the CPS network.
Since the tasks of the CPS network are scheduled by the system management, the system management know what application is running on the CPS network at any time.
Once the network resource usage is compromised, the recognized application label would be different from expected label by the system management. 
Assume the recognition accuracy of the trained $k$-NN model for application $A_{p} (p\in[1,N_{A}])$ is $R_{p}$, we set the accuracy threshold $Th_{p}$ of it as $Th_{p}=R_{p}*\theta_{th}$.
For application $A_p$ in period of anomaly detection, if the recognition accuracy $R^{'}_{p}$ is less than $Th_{p}$, we consider the network resource usage of the CPS network as abnormal and otherwise, we consider it as normal.

% \begin{figure}%[H]
% \centering
% \includegraphics[width=0.35\textwidth]{figures/IoT_System_Testing.pdf}
% \caption{Flow diagram of detecting anomalies}
% \label{fig:IoT_System_Testing}
% \end{figure}

\section{Experiment Results}
\label{sec:ExperimentResultsAndDiscussion}
In this section, we study four real-world applications, aggregation, broadcast, consensus and collection. 
In CPS network, these four decentralized applications are very popular and they have different communication styles. 
In aggregation, each leaf device sends data to its father devices.
The father devices process the received data and send same-size data to their father.
As to broadcast, starting with root device, data are sent from father devices to their children devices. 
For consensus, each device sends data to its neighbors and calculate received data and send data and calculate received data again and again till convergence is reached. 
The Distributed Gradient Descent (DGD) is very similar to consensus.
Each device do certain number of iterations of sending data to its neighbor and calculating received data.
The difference is for DGD, each device is synchronized with its neighboring devices, but for consensus, each device is asynchronous with its neighboring devices.

The experiments are based on CORE Simulator.
CORE Simulator is a network emulation tool that uses virtualization provided by FreeBSD jails, Linux OpenVZ containers, and Linux namespaces containers \cite{Ahrenholz2010ComparisonPlatforms}.
We set up 100 devices on it, as shown in Fig \ref{fig:IoT_System_100Nodes}.
We run every application on CORE Simulator for 200 times and the network throughput are recorded every second. 
We also run the CORE Simulator with executing none application as base reference for 200 times and record the network throughput every second as well.
The normalization range $L$ is empirically set as 9.
In order to shorten the training time of $k$-NN model, we use principal component analysis to reduce the dimensions of the feature data.
We choose to preserve 95\% of the information of the communication patterns, reducing the original 210 dimensions to only 11 dimensions.
The running time of these applications and the average network throughput of every CPS device throughput the running time of every application are shown as in Table \ref{tab:The_information_of_application}.
%\vspace{-1cm}

\begin{table}[H]
    \centering
    \caption{The information of applications.}
    \label{tab:The_information_of_application}
    \scalebox{1.1}{
        \begin{tabular}{|c|c|c|c|c|c|c|c|c|c|}
          \hline
          Application         & T (s)          & Average network throughput (kbps)\\ \hline
          Aggregation         & 195            & 44.25         \\ \hline
          Broadcast           & 195            & 23.77         \\ \hline
          Consensus           & 202            & 499.04        \\ \hline
          DGD                 & 242            & 533.35        \\ \hline
          Base reference      & 300            & 10.20         \\ \hline
        \end{tabular}
    }
\end{table}
%\vspace{-1.5cm}
In our experiment, we randomly choose 80\% of the communication patterns of every application as training data and label them respectively.
The rest 20\% of the communication patterns of every application are used to test the accuracy of the proposed recognition method.

\subsection{Accuracy of proposed method}
\begin{figure}%[H]
\centering
\includegraphics[width=0.48\textwidth]{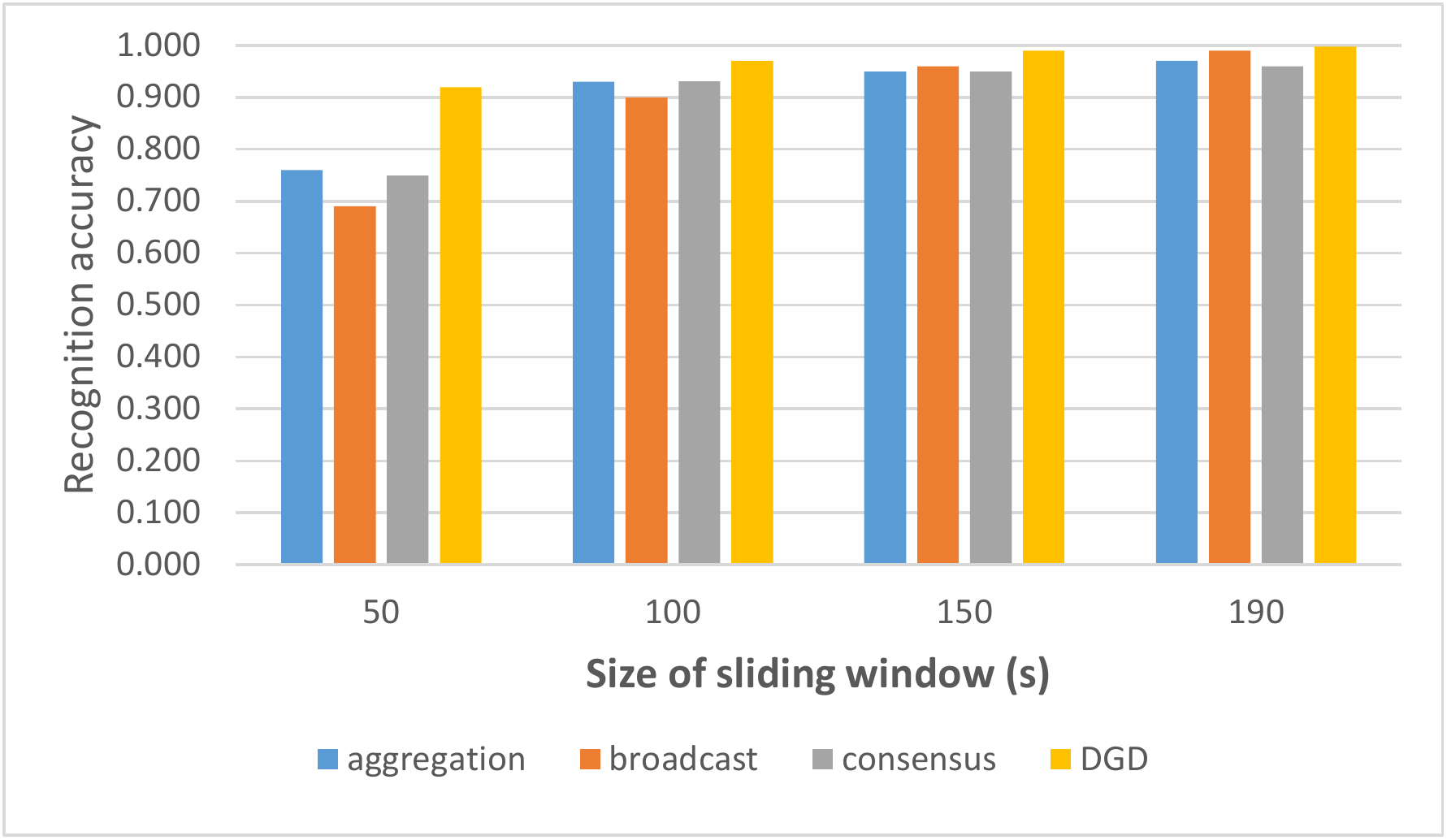}
%\vspace{-0.3cm}
\caption{The recognition accuracy of different sizes of the sliding window}
\label{fig:IoT_System_Characterizing_ErrorRate}
\end{figure}

This section presents the accuracy of the proposed method charactering the applications running on the CPS network. 
As shown in Fig. \ref{fig:IoT_System_Characterizing_ErrorRate}, the horizontal axis strands for the size of the sliding window $T_{S}$ and the vertical axis stands for recognition accuracy.
As we can see, the proposed method has achieved high accuracy.
When $T_{S}$ is 50s, the recognition accuracy of the broadcast is 68\%, which is lowest among all the applications.
With the $T_{S}$ increasing, the recognition accuracy of all the applications become higher.
When $T_{S}$ increase to 190s \footnote{Note that we don't evaluate the sliding window $T_{S}$ being 200s because the running time of aggregation and broadcast are 195s, less than 200s.}, the recognition accuracy of all the applications are higher than 95\%.
However, $T_{S}$ also controls the granularity of the communication patterns.
Smaller communication patterns could more details of the feature of the network resource usage. 
Hence, we have to compromise between the recognition accuracy and the granularity of communication patterns.

%\commentFL{legends are too small}

\subsection{Anomaly detection}
%\commentWS{Figure 10 does not show necessary information about the anomaly detection.}
In this section, we show our proposed method is capable of detecting abnormal network traffic.
As mentioned in section \ref{sec:attack_model}, both of DoS attack and botnet attack could bring in extract network traffic to almost every CPS device.
Thus we simulate the anomaly network traffic by adding different intensity of random network throughput to every device when the CPS devices are running.
We choose anomaly network throughput that equal to 10\%, 20\% and 30\% of the normal average throughput for every application.
%The average network throughput of the applications is shown in Table \ref{tab:The_information_of_application} 
For each intensity of anomaly network throughput, We run every application for 40 times.
We set $T_{S}$ as 100s since the trained $k$-NN model archives good balance between the recognition accuracy and the granularity of communication patterns when the $T_{S}$ is 100s.
We set the threshold parameter $\theta_{th}$ as 95\%.
The results are shown as in Fig. \ref{tab:IoT_System_Characterzing_ErrorRate_with_Anomaly}.

\begin{table}[H]
    \centering
    \caption{Comparison between threshold recognition accuracy and recognition accuracy with anomaly network resource usage}
    \label{tab:IoT_System_Characterzing_ErrorRate_with_Anomaly}
    \scalebox{0.99}{
        \begin{tabular}{|c|c|c|c|c|c|c|c|c|c|}
          \hline
                            & Aggregation  & Broadcast  & Consensus  & DGD    \\ \hline
          Normal            & 0.93         & 0.90       & 0.93       & 0.97   \\ \hline
          Thresholds        & 0.88         & 0.86       & 0.88       & 0.92   \\ \hline
          With 10\% anomaly & 0.78         & 0.75       & 0.53       & 0.35   \\ \hline
          With 20\% anomaly & 0.60         & 0.50       & 0.40       & 0.30   \\ \hline
          With 30\% anomaly & 0.46         & 0.25       & 0.32       & 0.27   \\ \hline
        \end{tabular}
    }
\end{table}

As we can see, for each intensity of anomaly network traffic, the recognition accuracy of aggregation, broadcast, Consensus and DGD is evidently lower than the corresponding thresholds, i.e., the proposed detection method succeeds in detecting the anomaly network traffic that caused by DoS attack or botnet attack.

\section{Conclusions}
\label{sec:conclusion}
This paper proposes to develop a network phenotyping mechanism based on network resource usage patterns and identify compromised network traffic. 
The proposed phenotyping method is based on transforming the network resource usage into time-series texture feature values, thus the computing complexity is reduced. 
We also apply both spatial and temporal analysis of the network resource usage.
The phenotyping method achieves every high accuracy of recognizing the communication patterns of four real-world decentralized applications.
Additionally, the paper proposes a detection method to verify whether the network resource usage of the CPS network is normal.
The detection method is straightforward and efficient.
In the future, we plan to combine the network resource usage with other resource usage to better phenotype the CPS network.
Besides, we only consider one application is running on the CPS network at a time, which is not always the case.
Hence, we also plan to characterize the CPS network that multiple applications run on it at the same time.

\ifCLASSOPTIONcaptionsoff
  \newpage
\fi

%% references section
\bibliographystyle{IEEEtran}
\bibliography{Mendeley.bib}{}

\begin{thebibliography}{10}
\providecommand{\url}[1]{#1}
\csname url@samestyle\endcsname
\providecommand{\newblock}{\relax}
\providecommand{\bibinfo}[2]{#2}
\providecommand{\BIBentrySTDinterwordspacing}{\spaceskip=0pt\relax}
\providecommand{\BIBentryALTinterwordstretchfactor}{4}
\providecommand{\BIBentryALTinterwordspacing}{\spaceskip=\fontdimen2\font plus
\BIBentryALTinterwordstretchfactor\fontdimen3\font minus
  \fontdimen4\font\relax}
\providecommand{\BIBforeignlanguage}[2]{{%
\expandafter\ifx\csname l@#1\endcsname\relax
\typeout{** WARNING: IEEEtran.bst: No hyphenation pattern has been}%
\typeout{** loaded for the language `#1'. Using the pattern for}%
\typeout{** the default language instead.}%
\else
\language=\csname l@#1\endcsname
\fi
#2}}
\providecommand{\BIBdecl}{\relax}
\BIBdecl

\bibitem{2017Gartner2016}
\BIBentryALTinterwordspacing
``{Gartner Says 8.4 Billion Connected "Things" Will Be in Use in 2017, Up 31
  Percent From 2016},'' 2017. [Online]. Available:
  \url{http://www.gartner.com/newsroom/id/3598917}
\BIBentrySTDinterwordspacing

\bibitem{Miorandi2012InternetChallenges}
\BIBentryALTinterwordspacing
D.~Miorandi, S.~Sicari, F.~De~Pellegrini, and I.~Chlamtac, ``{Internet of
  things: Vision, applications and research challenges},'' \emph{Ad Hoc
  Networks}, vol.~10, no.~7, pp. 1497--1516, 2012. [Online]. Available:
  \url{http://dx.doi.org/10.1016/j.adhoc.2012.02.016}
\BIBentrySTDinterwordspacing

\bibitem{Arias2017SecurityEra}
O.~Arias, K.~Ly, and Y.~Jin, ``{Security and Privacy in IoT Era},'' \emph{Smart
  Sensors at the IoT Frontier}, pp. 351--378, 2017.

\bibitem{Langner2011Stuxnet:Weapon}
R.~Langner, ``{Stuxnet: Dissecting a cyberwarfare weapon},'' \emph{IEEE
  Security and Privacy}, vol.~9, no.~3, pp. 49--51, 2011.

\bibitem{BencsathB.PekG.ButtyanL.andFelegyhazi2011Duqu:Wild}
P.~G. B.~L. Bencs{\'{a}}th, B. and M.~F{\'{e}}legyh{\'{a}}zi, ``{Duqu: A
  Stuxnet-like malware found in the wild},'' \emph{CrySyS Lab Technical
  Report}, vol.~14, pp. 1--60, 2011.

\bibitem{Seshadri2005Pioneer:Systems}
A.~Seshadri, A.~Perrig, M.~Luk, L.~Van~Doom, E.~Shi, and P.~Khosla, ``{Pioneer:
  Verifying code integrity and enforcing untampered code execution on legacy
  systems},'' \emph{Operating Systems Review (ACM)}, vol.~39, no.~5, pp. 1--16,
  2005.

\bibitem{Group-Trusted-Computing2014TPMSpecifications}
\BIBentryALTinterwordspacing
{Group-Trusted-Computing}, ``{TPM Library Specification (TPM 2.0
  specifications)},'' 2014. [Online]. Available:
  \url{https://trustedcomputinggroup.org/tpm-library-specification/}
\BIBentrySTDinterwordspacing

\bibitem{Winter2008TrustedPlatforms}
J.~Winter, ``{Trusted Computing Building Blocks for Embedded Linux-based ARM
  Trustzone Platforms},'' \emph{Proceedings of the 3rd ACM Workshop on Scalable
  Trusted Computing}, pp. 21--30, 2008.

\bibitem{Francillon2014AAttestation}
\BIBentryALTinterwordspacing
A.~Francillon, Q.~Nguyen, K.~B. Rasmussen, and G.~Tsudik, ``{A minimalist
  approach to Remote Attestation},'' in \emph{Design, Automation {\&} Test in
  Europe Conference {\&} Exhibition (DATE), 2014}.\hskip 1em plus 0.5em minus
  0.4em\relax New Jersey: IEEE Conference Publications, 2014, pp. 1--6.
  [Online]. Available:
  \url{http://ieeexplore.ieee.org/xpl/articleDetails.jsp?arnumber=6800458}
\BIBentrySTDinterwordspacing

\bibitem{EldefrawyKarimandTsudikGeneandFrancillonAurelienandPerito2012SMARTRust}
K.~Eldefrawy, G.~Tsudik, A.~Francillon, and D.~Perito, ``{SMART : S ecure and M
  inimal A rchitecture for ( Establishing a Dynamic ) R oot of T rust},''
  \emph{In Network and Distributed System Security Symposium}, vol.~12, pp.
  1--15, 2012.

\bibitem{Koeberl2014TrustLite:Devices}
\BIBentryALTinterwordspacing
P.~Koeberl, S.~Schulz, A.-R. Sadeghi, and V.~Varadharajan, ``{TrustLite: A
  Security Architecture for Tiny Embedded Devices},'' \emph{Proceedings of the
  European Conference on Computer Systems (EuroSys)}, pp. 1--14, 2014.
  [Online]. Available: \url{http://dl.acm.org/citation.cfm?id=2592798.2592824}
\BIBentrySTDinterwordspacing

\bibitem{Brasser2015TyTAN:Devices}
\BIBentryALTinterwordspacing
F.~Brasser, B.~El~Mahjoub, A.-R. Sadeghi, C.~Wachsmann, and P.~Koeberl,
  ``{TyTAN: Tiny trust anchor for tiny devices},'' \emph{Proceedings of the
  52nd Annual Design Automation Conference on - DAC '15}, pp. 1--6, 2015.
  [Online]. Available:
  \url{http://dl.acm.org/citation.cfm?doid=2744769.2744922}
\BIBentrySTDinterwordspacing

\bibitem{Liu2009VirusMeter:Spies}
L.~Liu, G.~Yan, X.~Zhang, and S.~Chen, ``{VirusMeter: Preventing Your Cellphone
  from Spies},'' in \emph{Recent Advances in Intrusion Detection: 12th
  International Symposium, RAID 2009, Saint-Malo, France, September 23-25,
  2009. Proceedings}, 2009, pp. 244--264.

\bibitem{Kang2011UsageSmartphones}
J.~M. Kang, S.~S. Seo, and J.~W.~K. Hong, ``{Usage pattern analysis of
  smartphones},'' in \emph{APNOMS 2011 - 13th Asia-Pacific Network Operations
  and Management Symposium: Managing Clouds, Smart Networks and Services, Final
  Program}, 2011.

\bibitem{Evans2014ComprehensiveStats}
T.~Evans, W.~L. Barth, J.~C. Browne, R.~L. Deleon, T.~R. Furlani, S.~M. Gallo,
  M.~D. Jones, and A.~K. Patra, ``{Comprehensive resource use monitoring for
  HPC systems with TACC stats},'' in \emph{Proceedings of HUST 2014: 1st
  International Workshop on HPC User Support Tools - Held in Conjunction with
  SC 2014: The International Conference for High Performance Computing,
  Networking, Storage and Analysis}, 2014.

\bibitem{Sorkunlu2017TrackingData}
\BIBentryALTinterwordspacing
N.~Sorkunlu, V.~Chandola, and A.~Patra, ``{Tracking System Behaviour from
  Resource Usage Data},'' \emph{arXiv preprint}, 5 2017. [Online]. Available:
  \url{http://arxiv.org/abs/1705.10756}
\BIBentrySTDinterwordspacing

\bibitem{Caviglione2016SeeingIntelligence}
L.~Caviglione, M.~Gaggero, J.~F. Lalande, W.~Mazurczyk, and M.~Urba{\'{n}}ski,
  ``{Seeing the unseen: Revealing mobile malware hidden communications via
  energy consumption and artificial intelligence},'' \emph{IEEE Transactions on
  Information Forensics and Security}, 2016.

\bibitem{Stavrou2017DDoSIoT}
A.~Stavrou, J.~Voas, and I.~Fellow, ``{DDoS in the IoT},'' \emph{Computer},
  vol.~50, pp. 80--84, 2017.

\bibitem{2002TextureExtraction}
\BIBentryALTinterwordspacing
``{Texture Feature Extraction},'' in \emph{Perspectives on Content-Based
  Multimedia Systems}, ser. The Information Retrieval Series.\hskip 1em plus
  0.5em minus 0.4em\relax Boston: Springer US, 2002, vol.~9, pp. 69--91.
  [Online]. Available: \url{http://link.springer.com/10.1007/b116171}
\BIBentrySTDinterwordspacing

\bibitem{Materka1998TextureReview}
A.~Materka and M.~Strzelecki, ``{Texture Analysis Methods – A Review},''
  \emph{Methods}, vol.~11, pp. 1--33, 1998.

\bibitem{Zhao2016CharacterizingMaps}
T.~Zhao, J.~Zhang, F.~Li, and K.~J. Marfurt, ``{Characterizing a turbidite
  system in Canterbury Basin, New Zealand, using seismic attributes and
  distance-preserving self-organizing maps},'' \emph{Interpretation}, vol.~4,
  no.~1, pp. SB79--SB89, 2 2016.

\bibitem{Baraldi1995InvestigationParameters}
A.~Baraldi and F.~Parmiggiani, ``{Investigation of the textural characteristics
  associated with gray level cooccurrence matrix statistical parameters},''
  \emph{IEEE Transactions on Geoscience and Remote Sensing}, vol.~33, no.~2,
  pp. 293--304, 1995.

\bibitem{Haralick1973TexturalClassification}
\BIBentryALTinterwordspacing
R.~M. Haralick, K.~Shanmugam, and I.~Dinstein, ``{Textural Features for Image
  Classification},'' \emph{IEEE Transactions on Systems, Man, and Cybernetics},
  vol. SMC-3, no.~6, pp. 610--621, 1973. [Online]. Available:
  \url{http://ieeexplore.ieee.org/document/4309314/}
\BIBentrySTDinterwordspacing

\bibitem{Benesty2009PearsonCoefficient}
J.~Benesty, J.~Chen, Y.~Huang, and I.~Cohen, ``{Pearson Correlation
  Coefficient},'' in \emph{Noise Reduction in Speech Processing}.\hskip 1em
  plus 0.5em minus 0.4em\relax Springer Berlin Heidelberg, 2009, pp. 1--4.

\bibitem{Altman1992AnRegression}
\BIBentryALTinterwordspacing
N.~S. Altman, ``{An Introduction to Kernel and Nearest-Neighbor Nonparametric
  Regression},'' \emph{The American Statistician}, vol.~46, no.~3, pp.
  175--185, 8 1992. [Online]. Available:
  \url{http://www.tandfonline.com/doi/abs/10.1080/00031305.1992.10475879}
\BIBentrySTDinterwordspacing

\bibitem{Ahrenholz2010ComparisonPlatforms}
J.~Ahrenholz, ``{Comparison of CORE network emulation platforms},''
  \emph{Proceedings - IEEE Military Communications Conference MILCOM}, pp.
  166--171, 2010.

\end{thebibliography}

%% biography section
\begin{IEEEbiography}[{\includegraphics[width=1in,height=1.25in,clip,keepaspectratio]{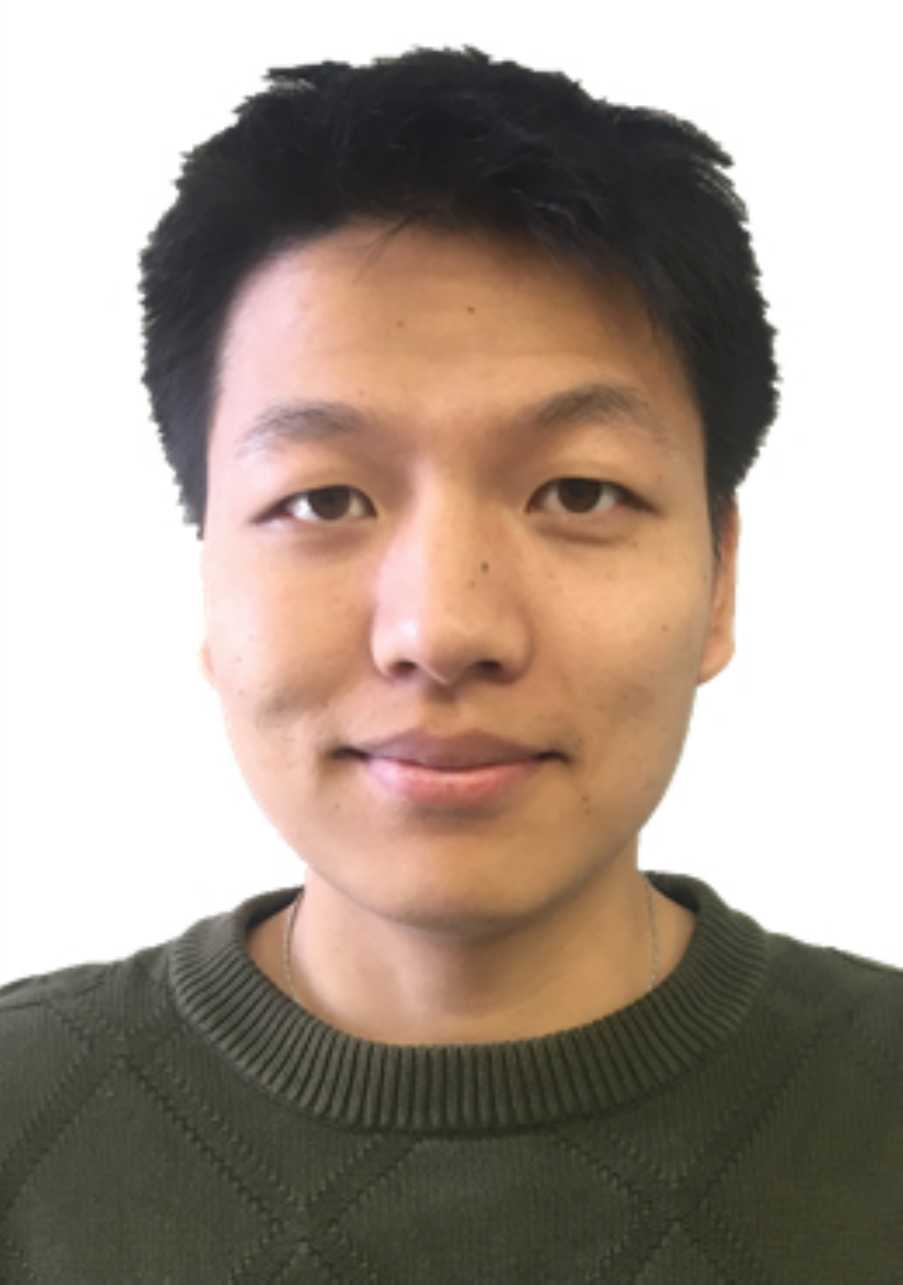}}]
{Minhui Zou} received the B.S. degree in computer science and technology from Chongqing University, China, in 2013.
Currently he is a Ph.D. student majoring in computer science and technology of the College of Computer Science, Chongqing University.
His current research interests include security of cryptographic system and side-channel attacks.
\end{IEEEbiography}

\begin{IEEEbiography}[{\includegraphics[width=1in,height=1.25in,clip,keepaspectratio]{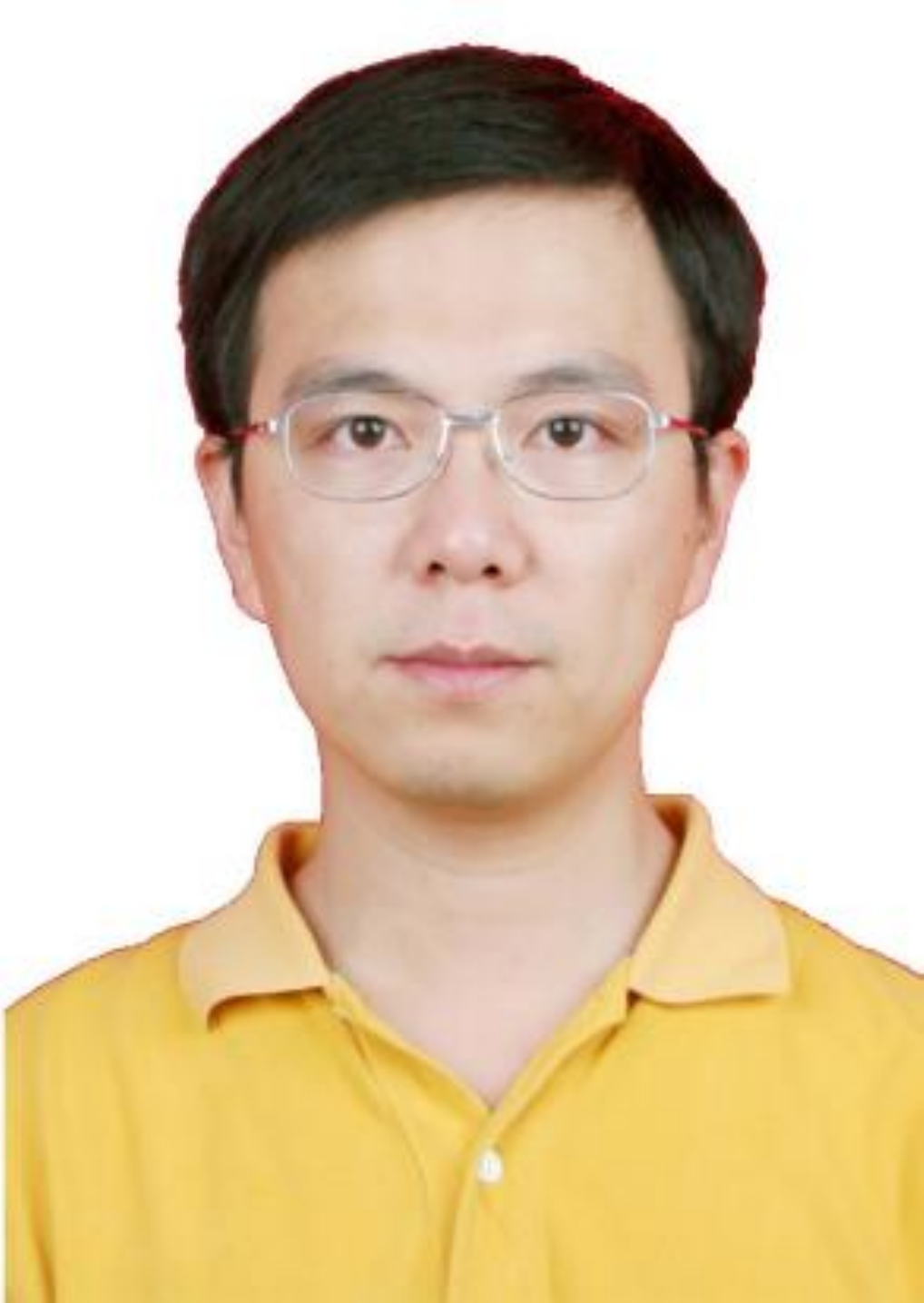}}]
{Chengliang~Wang} received his B.S. degree in mechatronics in 1996, the M.S. degree in precision instruments and machinery in 1999, and the Ph.D. degree in control theory and engineering in 2004, all from Chongqing University, China.
He is now a professor of Chongqing University.
His research interests include smart control for complex system, the theory and application of artificial intelligence, wireless network and RFID research.
He is a senior member of China computer science association and member of America Association of Computing Machinery.
\end{IEEEbiography}

\begin{IEEEbiography}
[{\includegraphics[width=1in,height=1.25in,clip,keepaspectratio]{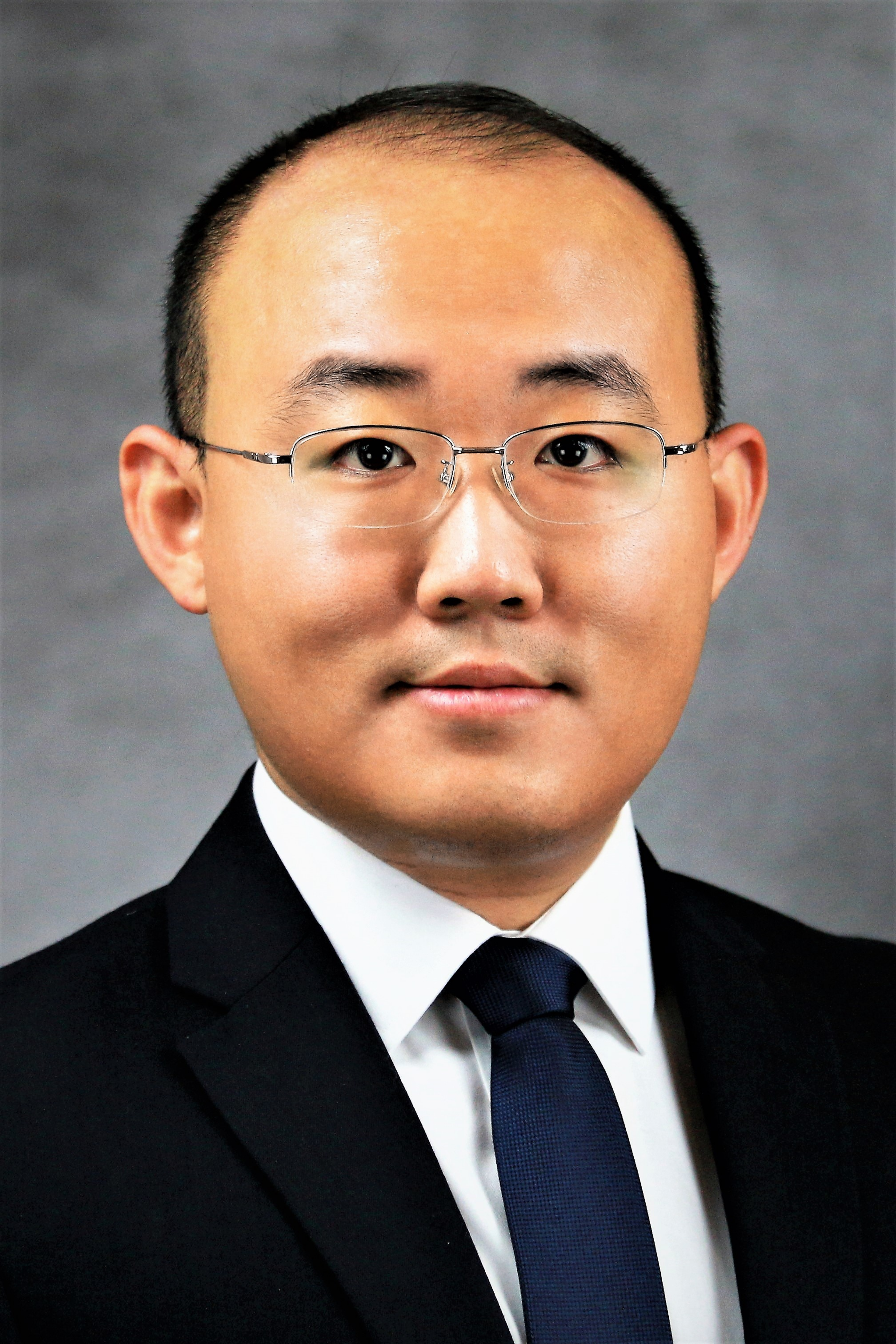}}]
{Fangyu Li} is a postdoctoral research associate in the College of Engineering, University of Georgia. He received his PhD in Geophysics from University of Oklahoma in 2017. His Master and Bachelor degrees are both in Electrical Engineering, obtained from Tsinghua University and Baihang University, respectively. His research interests include signal processing, seismic imaging, geophysical interpretation, machine learning and distributed system.
\end{IEEEbiography}

\begin{IEEEbiography}[{\includegraphics[width=1in,height=1.25in,clip,keepaspectratio]{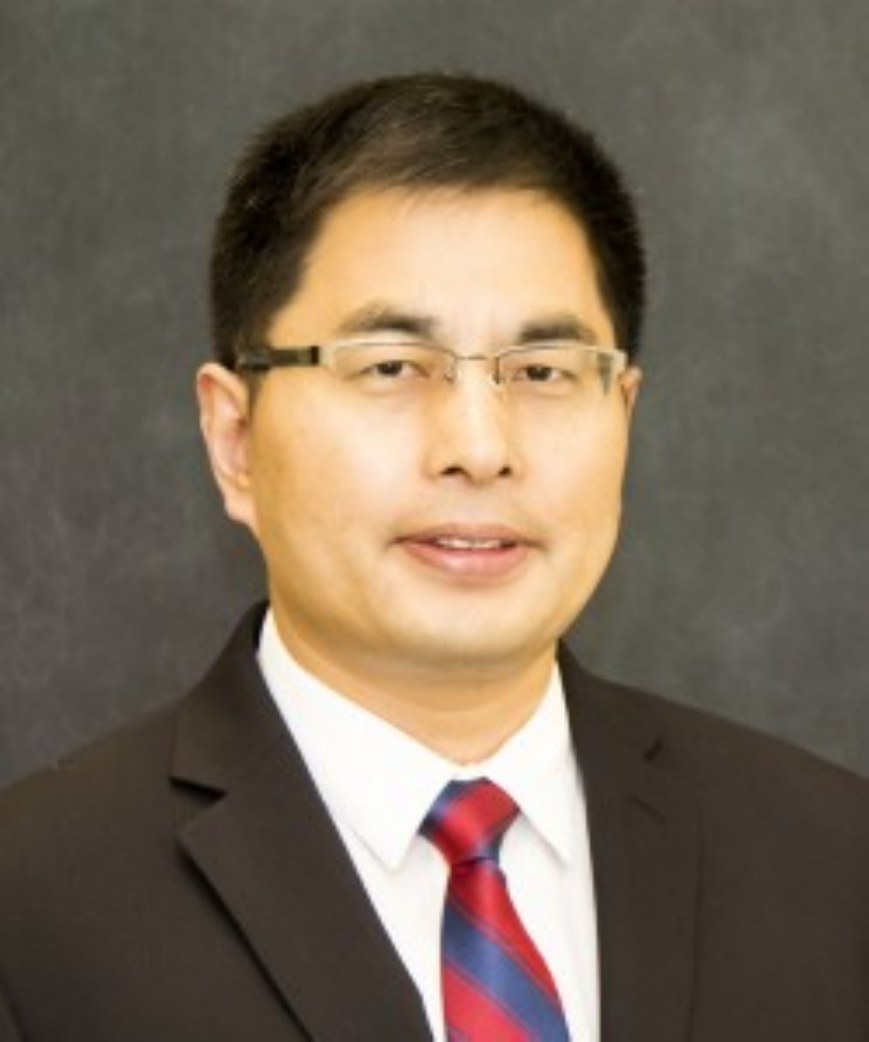}}]
{WenZhan~Song} is now Georgia Power Mickey A. Brown Professor in College of Engineering, University of Georgia. His research mainly focuses on sensor web, smart grid and smart environment where sensing, computing, communication and control play a critical role and need a transformative study. His research has received 6 million+ research funding from NSF, NASA, USGS, Boeing and etc since 2005.
\end{IEEEbiography}

\end{document}